\documentclass[print]{nst}

\usepackage{subfigure,dcolumn}
\usepackage[T2A,T1]{fontenc}
\usepackage[russian,english]{babel}
\usepackage{threeparttable}
\usepackage[utf8]{inputenc}
\usepackage[T1]{fontenc}
\usepackage{mathptmx}
\usepackage{booktabs}
\usepackage{textcomp}
\usepackage{amsmath}
\usepackage{multirow}
\usepackage{epstopdf}
\usepackage{jabbrv}
\usepackage{listings}
\begin{document}

\title{Rapid interrogation of special nuclear materials by  combining scattering and transmission nuclear resonance fluorescence spectroscopy}\thanks{This work was supported by the National Natural Science Foundation of
China (Grant No. 11675075); Youth Talent Project of Hunan Province, China (Grant No. 2018RS3096); Independent Research Project of Key Laboratory of Plasma Physics, CAEP (Grant No. JCKYS2020212006) and Innovation and Entrepreneurship Training Program for College Students of University of South China (Grant No. X2019083).}

\author{Haoyang Lan}
\affiliation{School of Nuclear Science and Technology, University of South China, Hengyang 421001, China}
\author{Tan Song}
\affiliation{School of Nuclear Science and Technology, University of South China, Hengyang 421001, China}
\author{Jialin Zhang}
\affiliation{School of Nuclear Science and Technology, University of South China, Hengyang 421001, China}
\author{Jianliang Zhou}
\affiliation{School of Nuclear Science and Technology, University of South China, Hengyang 421001, China}
\author{Wen Luo}
\email[Corresponding author, ]{Wen Luo, wenluo-ok@163.com}
\affiliation{School of Nuclear Science and Technology, University of South China, Hengyang 421001, China}
\affiliation{National Exemplary Base for International Sci \& Tech. Collaboration of Nuclear Energy and Nuclear Safety, University of South China, Hengyang 421001, China}

\begin{abstract}
The smuggling of special nuclear materials (SNMs) across national borders is becoming a serious threat to nuclear nonproliferation. This paper presents a feasibility study on the rapid interrogation of concealed SNMs by combining scattering and transmission nuclear resonance fluorescence (sNRF and tNRF) spectroscopy. In sNRF spectroscopy, SNMs such as $^{235, 238}$U are excited by a wide-band photon beam of appropriate energy and exhibit unique NRF signatures. Monte Carlo simulations show that one-dimensional scans can realize isotopic identification of concealed $^{235, 238}$U when the detector array used for interrogation has sufficiently high energy resolution. The simulated isotopic ratio $^{235}U/^{238}U$ is in good agreement with the theoretical value when the SNMs are enclosed in relatively thin iron. This interrogation is followed by tNRF spectroscopy using a narrow-band photon beam with the goal of obtaining tomographic images of the concealed SNMs. The reconstructed image clearly reveals the position of the isotope $^{235}$U inside an iron rod. It is shown that the interrogation time of sNRF and tNRF spectroscopy is one order of magnitude lower than that when only tNRF spectroscopy is used and results in a missed-detection rate of 10$^{-3}$. The proposed method can also be applied for isotopic imaging of other SNMs such as $^{239, 240}$Pu and $^{237}$Np.
\end{abstract}
\maketitle
\thispagestyle{empty}

\section{Introduction}
The smuggling of special nuclear materials (SNMs) across borders and through ports of entry is one of the greatest threats to global security. The Incident Trafficking Database, which was developed by the International Atomic Energy Agency to record incidents of illicit trafficking in nuclear and other radioactive materials, was notified of several hundred incidents that involved the deliberate trafficking or malicious
use of certain nuclear and radioactive materials~\cite{iaeainicdents}. Previous studies have illustrated how these materials, if obtained
in sufficient quantities by actors such as terrorist groups, could cause significant death, destruction, and disruption~\cite{ferguson2005four}. To reduce this threat to homeland security, efforts have been made to develop accurate, effective, and practical ways to interrogate SNMs, especially uranium and plutonium.

Passive detection systems, which exploit the $\gamma$ rays and/or neutrons naturally emitted from radioactive isotopes, can be easily deployed to identify SNMs by delivering a low radiation dose to the inspected target~\cite{kouzes2008passive,cester2012special}. However, this detection method may be inapplicable when the interrogated object is shielded, because  the intensity and energy of the spontaneous radiation are fairly low in most cases. Therefore, the inspection of SNMs requires active detection techniques that utilize external radiation sources such as muons~\cite{thomay2013binned,guardincerri2015detecting,baesso2012high,pan2019experimental}, neutrons~\cite{slaughter2003detection,norman2003signatures,huang2019study,paff2014gamma,huang2019element}, and photons~\cite{mueller2014novel,zier2014high,henderson2018experimental}.
However, active interrogation systems using cosmic-ray muons generally require long data acquisition times and large detection systems, and those using photon-/neutron-induced fission face measurement challenges arising from the high background of intense interrogating radiation.

Recently, nondestructive detection methods based on nuclear resonance fluorescence (NRF) have been proposed in the context of industrial applications~\cite{Lan2021,Beck1998,Hajima2009,Habs2011,yu2019ultrafast,hayakawa2009nondestructive,toyokawa2011nondestructive} as well as nuclear safeguards~\cite{zkproof1,zkproof2}. NRF is the process of resonant excitation of nuclear levels of an isotope of interest by the absorption of electromagnetic radiation and subsequent decay of these levels by photon emission. Because the resonant energies are unique to an isotope, the emitted photons can be used as signatures for isotope identification. In addition, $\gamma$-ray beams generated by laser Compton scattering (LCS), which have been used for research on nuclear physics~\cite{Utsunomiya2019,Ur2016} and nuclear astrophysics~\cite{Lan2018} as well as industrial and medical applications~\cite{Zhu2016,Irani2014,luo2016estimates,luo2016data,luo2016production,artun2018investigation}, have excellent characteristics such as good directivity, a narrow-band spectrum, energy tunability, and moderate/high intensity. Owing to these unique features, the LCS $\gamma$-ray beam is regarded as a good candidate to excite NRF and thus to interrogate SNMs. Previous studies~\cite{daito2016simulation,SULIMAN2016,zen2019demonstration,ali2020selective} have proposed an effective method, namely, transmission-NRF-based computed tomography (tNRF-CT), for tomographic imaging of high-density and high-$Z$ objects.

However, tNRF-CT relies on a narrow-band beam with suitable energy for accurate evaluation of the attenuation factors associated with both atomic processes and NRF interactions. Without prior isotope identification, it seems difficult and time-consuming to interrogate SNMs with multiple nuclei and isotopes by scanning the beam energy and thus checking for every suspicious nuclear species.
\begin{figure*}
\includegraphics[width=12cm,clip]{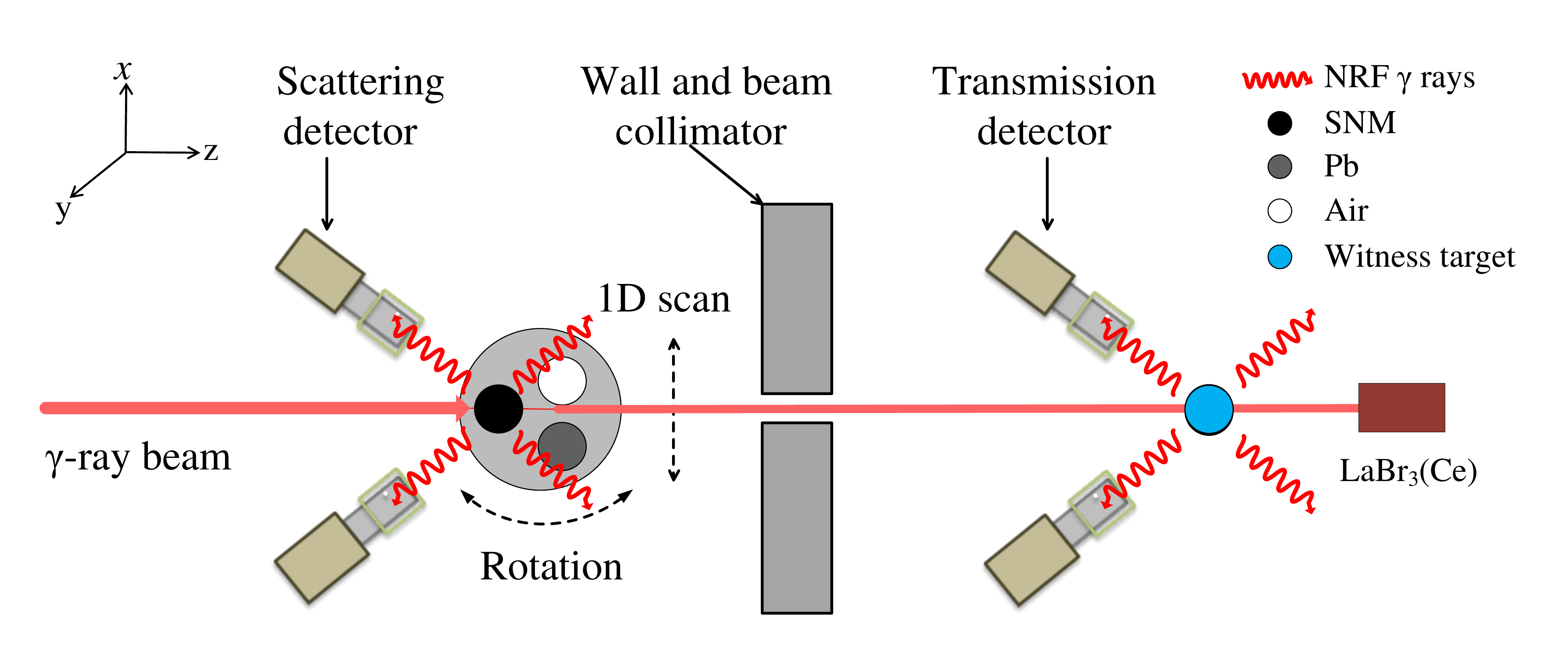}
\caption{\label{f1} Schematic illustration of SNM interrogation. The interrogated object consisted of three 10-mm-diameter rods made of uranium, lead, and air wrapped in a 30-mm-diameter iron cylinder. The NRF $\gamma$ rays scattered from the object and witness target were recorded by the scattering detectors in sNRF spectroscopy and the transmission detectors in tNRF spectroscopy, respectively. The $\gamma$ rays transmitted through the witness target were recorded by a LaBr$_{3}$(Ce) detector. A shielding wall was used to prevent the scattered $\gamma$ rays from entering the transmission detectors. }
\end{figure*}

In this paper, we propose combining scattering NRF (sNRF) and tNRF spectroscopy to rapidly realize isotope identification and tomographic imaging of SNMs such as $^{235, 238}$U. A schematic illustration of the proposed method is shown in Fig.~\ref{f1}.
In sNRF spectroscopy, a one-dimensional (1D) scan is performed using a wide-band $\gamma$-ray beam that covers exactly the principal resonant energies of $^{235, 238}$U. From the sNRF spectra, one can determine whether $^{235}$U and/or $^{238}$U is present in the interrogated object. Moreover, the sNRF yields can be used to deduce the isotopic ratio of $^{235}$U to $^{238}$U. We then perform tNRF spectroscopy on the isotope of interest ($^{235}$U or $^{238}$U), acquiring a CT image of the interrogated object using a narrow-band $\gamma$-ray beam covering exactly the resonant energy of a specific isotope. Simulations show that the presence of the $^{235,238}$U isotopes and the $^{235}U/^{238}U$ ratio are readily revealed by sNRF spectroscopy with high significance  in a reasonable time. The tNRF-CT technique provides a tomographic image of a $^{235}$U rod, lead rod, and air column wrapped in an iron shield. The combination of sNRF and tNRF spectroscopy can provide knowledge of not only the isotopic composition but also the spatial distribution of SNMs. The results show that it can shorten the interrogation time by one order of magnitude owing to the strong response of SNMs to sNRF spectroscopy. In addition, the feasibility of isotopic imaging of other SNMs ($^{239, 240}$Pu and $^{237}$Np) is discussed considering the attenuation factor of the on-resonance photon beam.

\section{Methods}
\subsection{NRF principle}

\begin{table*}
  \centering
  \caption{Resonant energy ($E_{r}$), width ($\Gamma$ or $\frac{g\Gamma_{0}^{2}}{\Gamma}$), and NRF cross section ($\sigma_{int}$) of $^{235}$U and $^{238}$U.}\label{tCS}
\begin{tabular*}{120mm}{@{\extracolsep{\fill}}ccccccc}
\toprule
SNM & $E_{r}$ (keV) & $\Gamma$ (meV) & $\frac{g\Gamma_{0}^{2}}{\Gamma}$ (meV) & $\sigma_{int}$ (eV$\cdot$b) & Zilges \cite{zilges1995strong} & Kwan \cite{kwan2011discrete} \\
\midrule
$^{235}$U    &1734	&	N/A   &17(3)   &  21.7(38)  &N/A &22(4)  \\
$^{235}$U    &1815	&	N/A   &7.7(9)  &   8.9(11)  &N/A &8.9(11)\\
$^{238}$U	 &1782	&13.8(17) &N/A 	&20.9(25)	& 21.9(25)& N/A\\
$^{238}$U	 &1793	&5.7(14)  &N/A	&4.6(12)	& 5.1(10) & N/A\\
$^{238}$U	 &1846	&14.7(19) &N/A	&21.8(28)   & 23.0(26)& N/A\\
\bottomrule
\end{tabular*}
\end{table*}

\begin{figure*}
\centering
\includegraphics[width=14cm,clip]{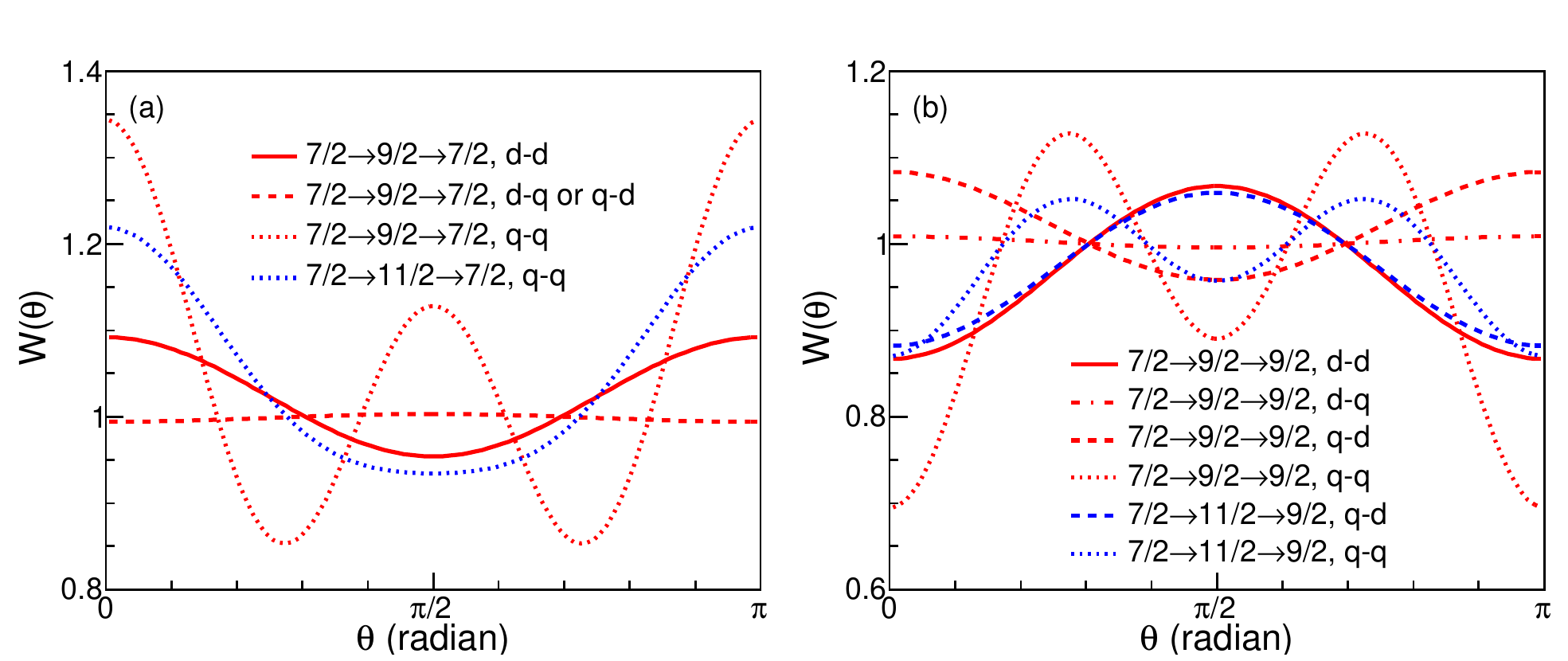}
\caption{\label{angular} $W(\theta)$ values for the NRF transition from the 1734 keV resonant state, which has hypothetical spin 9/2 (red) or 11/2 (blue), to the ground state with spin 7/2 (a) and to the first excited state with spin 9/2 (b). In panel (a), three possible multipolarity combinations are allowed: only dipolar  (solid lines), only quadrupolar (dotted lines), and one dipolar transition and one quadrupolar transition (dashed lines). In panel (b), four hypothetical multipolarity combinations are allowed: only dipolar (solid lines), only quadrupolar (dotted lines), dipolar-quadrupolar (dot-dashed lines), and quadrupolar-dipolar (dashed lines). }
\end{figure*}

The NRF cross section for absorption via the resonant energy level $E_{r}$ can be expressed by the Breit--Wigner distribution~\cite{Franz1959}:
\begin{eqnarray}
\centering
\sigma_{NRF}(E)=\frac{g}2\pi \frac{(\hbar c)^2}{E_{r}^2}\frac{\Gamma\Gamma_{0}}{(E-E_{r})^2+(\Gamma/2)^2},
\label{eqHF1}
\end{eqnarray}
\noindent
where $\Gamma$ is the width of the level at $E_{r}$, $\Gamma_{0}$ is the partial width for transitions between $E_{r}$ and the ground state, $\hbar$ is the Planck constant, and $c$ is speed of light.

In practice, the NRF cross section should be calculated taking into account Doppler broadening. If the true Voigt profile is approximated as a Gaussian profile, Eq.~\ref{eqHF1} then becomes~\cite{ogawa2016development}
 \begin{eqnarray}
 \sigma_{D}(E)\approx(\frac{\hbar c}{E_{r}})^{2}\frac{\pi^{3/2}}{\sqrt{2}\Delta}g \frac{\Gamma_{0}^{2}}{\Gamma}exp(\frac{(E-E_{r})^2}{2\Delta^2}),
 \label{eq4}
\end{eqnarray}
Here $\Delta$ = $E_{r}\sqrt{{k_{B}T}/{Mc^2}}$ is the Doppler width, $k_{B}$ is the Boltzmann constant, $T$ is the absolute temperature, and $M$ is the mass number of the nucleus. According to Eq.~\ref{eq4}, one can calculate the integrated NRF cross section $\sigma_{int}$ for $^{235}$U and $^{238}$U. As shown in Table~\ref{tCS}, the $\sigma_{int}$ values are consistent with experimental data for $^{235}$U and $^{238}$U~\cite{zilges1995strong,kwan2011discrete}. The NRF lines at 1734 keV ($^{235}$U) and 1782 keV ($^{238}$U) have NRF cross sections of 22.0 and 21.9 eV$\cdot$b, respectively. Considering their relatively large $\sigma_{int}$ values, these two separate NRF lines are priorities for the identification of uranium, which is selected as a typical SNM here.

Because of conservation of energy and momentum, a free nucleus undergoing NRF will recoil with kinetic energy $E_{rec}$, which is determined by the Compton-like formula
 \begin{equation}
 E_{rec} = E_{r} \left[1 - \frac{1}{1 + E_{r}(1 - cos \theta)/Mc^{2})}\right] \\
 \approx \frac{E_{r}^2}{2Mc^2} (1 - cos \theta),
\label{eq6}
 \end{equation}
where $\theta$ is the scattering angle of the photon relative to its incident direction.

NRF is generally considered to occur only between states that differ by two or fewer units of angular momentum. The angular distribution of NRF $\gamma$ rays is analogous to that of $\gamma$-ray cascades.
For an NRF interaction of transitions $J_{a}(L_{1})J_{b}(L_{2})J_{c}$, where $L_{1}$ and $L_{2}$ are the multipole orders of excitation and de-excitation, respectively, the angular distribution $W(\theta)$ can be written as~\cite{Fagg1959}
 \begin{eqnarray}
 W(\theta)=1+A_{2}P_{2}(cos\theta)+...+A_{2n}P_{2n}(cos\theta),
\label{eq7}
\end{eqnarray}
where $P_{2n}(cos\theta)$ is the Legendre polynomial expansion, and $A_{2n}$ is given by
 \begin{eqnarray}
 A_{2n}=F_{2n}(L_{1}J_{a}J_{b}) F_{2n}(L_{2}J_{c}J_{b}),
\label{eq8}
\end{eqnarray}
where $F_{2n}(L_{1}J_{a}J_{b})$ and $F_{2n}(L_{2}J_{c}J_{b})$ are constants that depend on the spin states of the transitions and photon multipolarities~\cite{Siegbahn1965}. For the resonant state at 1782 keV ($^{238}$U), the NRF follows a transition sequence of 0 $\rightarrow$ 1 $\rightarrow$ 0, whose angular correlation can be expressed as $W(\theta)=0.75\times(1+cos^{2}\theta)$. By contrast, at 1734 keV ($^{235}$U), $W(\theta)$ depends on the spin, $J$ = 9/2 or 11/2. Because this state can de-excite to the first excited state and ground state of $^{235}$U, several multipolarity combinations are obtained according to the spin selection rule (see Fig.~\ref{angular}). However, it is still impossible to obtain an exact expression of $W(\theta)$ because their mixing ratios remain unknown. For simplicity, we employ an isotropic $W(\theta)$ for NRF $\gamma$-ray emissions in the simulations. In fact, a non-isotropic angular distribution would contribute at most a $\sim$10\% fluctuation to the NRF yields in our configuration (see Fig.~\ref{f1}). More details are given in Sec.~\ref{sec-discussion}.

\subsection{Scattering NRF spectroscopy}
\label{sec-interrogation}

To realize SNM identification and isotope ratio prediction, 1D sNRF spectroscopy is applied. As shown in Fig.~\ref{f1}, a quasi-monochromatic $\gamma$-ray beam impinges on the target to be interrogated, causing resonant (NRF) and non-resonant (Compton scattering, pair production, and photoelectric absorption) interactions. The backscattered NRF $\gamma$ rays are measured by four high-purity germanium (HPGe) detectors (scattering detectors) located at 135$^{\circ}$ from the beam direction in order to take advantage of the decreasing intensity of non-resonantly backscattered radiation. The horizontal position ($x$) is varied from -15 to 15 mm in eight steps of 3.75 mm each. A total of eight sNRF $\gamma$-ray spectra are obtained.

In sNRF spectroscopy, one can use the NRF cross section $\sigma_{NRF}(E)$ and angular distribution $W(\theta)$ to construct a semi-analytical expression for the expected NRF counts.
For a photon beam of incident flux $I(E)$ interacting with the target, a small part of the photon flux near the resonant energy $E_{r}$ will undergo resonant (NRF) and non-resonant (atomic) interactions. The resulting NRF yield then produces a double-differential rate of NRF detections in the infinitesimal solid angle $d\Omega$,
\begin{equation}
\frac{d^{2}Y}{dE d\Omega}= I(E) \mu_{NRF}(E) \frac{W(\theta)}{4\pi} \frac{1 - exp\left[ - L \mu_{eff}(E,E') \right]}{\mu_{eff}(E,E')} \epsilon_{d}(E'),
\label{eq-Event}
\end{equation}
where $E$ and $E'$ are the energy of the incident photons and scattered NRF photons, respectively;
$L$ is the thickness of the irradiated target; $\epsilon(E')$ is the intrinsic photopeak detection efficiency; $\mu_{NRF}(E) = N\sigma_{NRF}(E)$ denotes the linear attenuation coefficient, with $N$ being the number density of interrogated isotopes; and $\mu_{eff}(E,E')$ is the effective attenuation coefficient, which is given by
\begin{eqnarray}
\mu_{eff}(E,E',\theta) = \mu_{NRF}(E) + \mu_{nr}(E) + \mu_{nr}(E').
\label{eq-mueff}
\end{eqnarray}
Here $\mu_{nr}(E)$ and $\mu_{nr}(E')$ are the non-resonant attenuation coefficients of the incident photons and NRF photons, respectively.

\subsection{Transmission NRF spectroscopy}
After sNRF spectroscopy is performed, a tNRF-CT technique is applied to perform tomographic imaging (see Fig.~\ref{f1}). The flux of the $\gamma$-ray beam transmitted through the target is preferentially attenuated (notched~\cite{Pruet2006Detecting}) around the resonant energy $E_{r}$ because the NRF cross section is much larger than those of the non-resonant interactions. This notched $\gamma$-ray beam further impinges on a witness target composed of suspicious isotopes so that the remaining $\gamma$ rays may undergo NRF in the witness target. Another array of four HPGe detectors (transmission detectors) are located at 135$^{\circ}$ to record the NRF photons produced at this stage. The resonant attenuation inside the interrogated object is then evaluated. The $\gamma$ rays transmitted through the witness target are diagnosed by a LaBr$_{3}$(Ce) detector to evaluate the non-resonant attenuation.
To obtain the CT images, the interrogated object is translated horizontally ($x$) from -15 to 15 mm (with a step length of 3.75 mm) and rotated (by $\theta_{r}$) from 0 to 180$^{\circ}$ (with a step length of 22.5$^{\circ}$); consequently, a total of 64 sets of spectra are obtained. In addition, a set of spectra without the interrogated object is obtained.

\begin{figure*}
\centering
\includegraphics[width=13cm,clip]{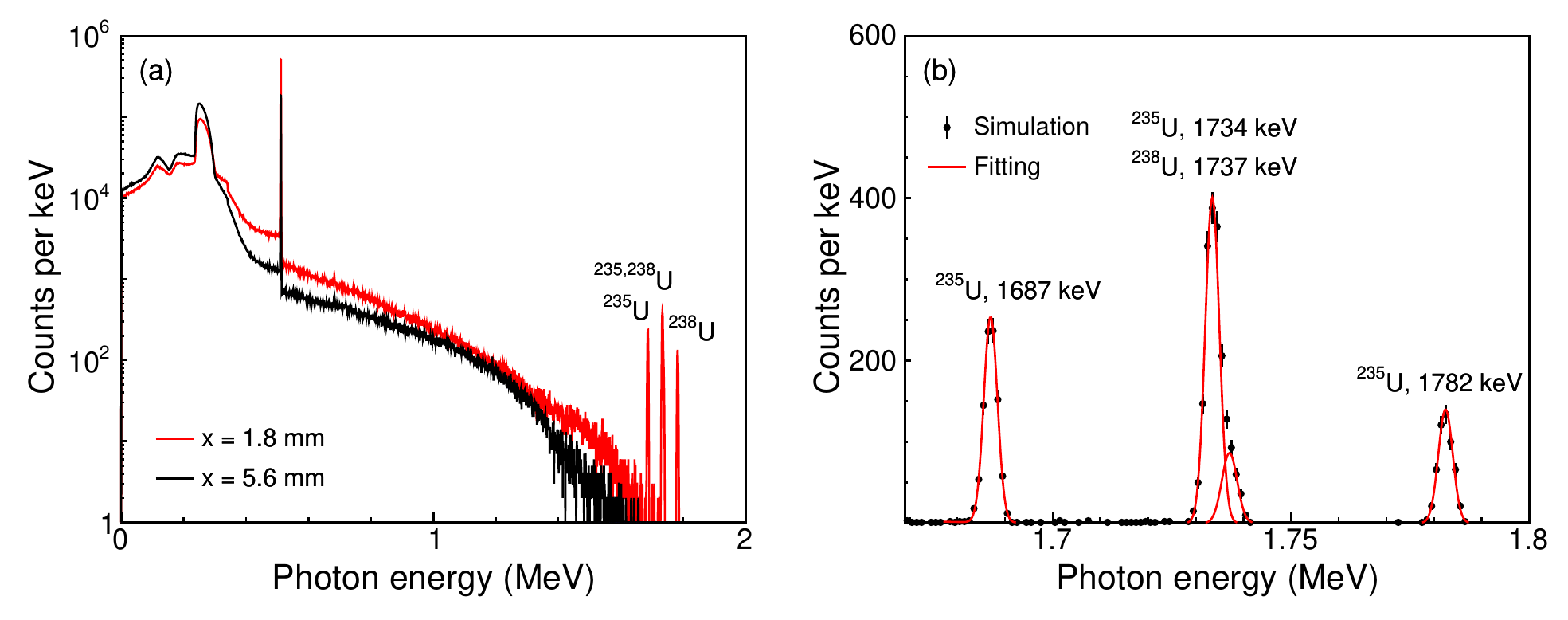}
\caption{\label{f5}(a) Simulated sNRF $\gamma$-ray spectra at scan points ($x$, $\theta_{r}$) = (1.8 mm, 0$^{\circ}$) (red line) and (5.6 mm, 0$^{\circ}$) (black line) recorded by scattering detectors located at 135$^{\circ}$. There are clear photopeaks at 1687, 1734, and 1782 keV. The density of the uranium rod is set to 19 g/cm$^{3}$. The isotopic composition of the uranium rod is set to 80\% $^{235}$U and 20\% $^{238}$U. Single and double escape peaks are not observed because the low-probability escapes are obscured by low-energy photons deposited in the detectors with large volume . (b) Magnified spectrum showing the NRF signal region of 1650-- 1800 keV. Solid lines represent the fitting curves of four Gaussian peaks plus an exponential background fit.}
\end{figure*}

\begin{table*}
  \centering
  \caption{The NRF yields, $Y_{1687}$, $Y_{1734,1737}$ and $Y_{1782}$, obtained with spectral peak fitting. $Y^{ded}_{1734}$ and $Y^{ded}_{1737}$ are extracted from $Y_{1687}$ and $Y_{1782}$ according to the branching ratios. The isotopic composition of the uranium rod is 80\% $^{235}$U and 20\% $^{238}$U.}
\begin{tabular*}{120mm}{@{\extracolsep{\fill}}ccccccc}
    \toprule
    \multirow{2}[4]{*}{$x$ (mm)} & \multicolumn{2}{c}{$^{235}$U} & \multicolumn{2}{c}{$^{238}$U} & \multirow{2}[4]{*}{$Y^{ded}_{1734,1737}$} & \multirow{2}[4]{*}{$Y_{1734,1737}$} \\
\cmidrule{2-3} \cmidrule{4-5}         & $Y_{1687}$ & $Y^{ded}_{1734}$ & $Y_{1782}$ & $Y^{ded}_{1737}$ &       &  \\
    \midrule
    1.8   & 920 $\pm$ 30 & 1518 $\pm$ 50 & 510 $\pm$ 23 & 281 $\pm$ 12 & 1798 $\pm$ 52 & 1825 $\pm$ 43 \\
    -1.8  & 918 $\pm$ 30 & 1514 $\pm$ 50 & 507 $\pm$ 23 & 279 $\pm$ 12 & 1793 $\pm$ 51 & 1821 $\pm$ 43 \\
    \bottomrule
    \end{tabular*}%
  \label{table-NRFyilds}%
\end{table*}%

\begin{table*}[htbp]
  \centering
  \caption{The sNRF yields and the expected isotope ratio $^{235}U/^{238}U$ for three isotopic compositions of $(^{235}U/^{238}U)_{theory}$ = 0.43, 1.00 and 4.00.}
\begin{tabular*}{120mm}{@{\extracolsep{\fill}}cccccc}
    \toprule
    \multirow{2}[2]{*}{$(^{235}U/^{238}U)_{theory}$} & \multirow{2}[2]{*}{$x$ (mm)} & \multirow{2}[2]{*}{$Y_{1687}$} & \multirow{2}[2]{*}{$Y_{1782}$} & \multirow{2}[2]{*}{$Y_{1734,1737}$} & \multirow{2}[2]{*}{$^{235}U/^{238}U$} \\
          &       &       &       &       &  \\
    \midrule
    \multirow{2}[4]{*}{0.43} & 1.8   & 376 $\pm$ 19 & 1634 $\pm$ 40 & 1500 $\pm$ 39 & 0.43 $\pm$ 0.09 \\
\cmidrule{2-6}          & -1.8  & 378 $\pm$ 19 & 1656 $\pm$ 41 & 1424 $\pm$ 38 & 0.42 $\pm$ 0.09 \\
    \midrule
    \multirow{2}[4]{*}{1.00} & 1.8   & 604 $\pm$ 25 & 1345 $\pm$ 37 & 1573 $\pm$ 40 & 0.83 $\pm$ 0.18 \\
\cmidrule{2-6}          & -1.8  & 579 $\pm$ 24 & 1278 $\pm$ 36 & 1626 $\pm$ 40 & 0.84 $\pm$ 0.19 \\
    \midrule
    \multirow{2}[4]{*}{4.00} & 1.8   & 920 $\pm$ 30 & 510 $\pm$ 23 & 1825 $\pm$ 43 & 3.28 $\pm$ 0.73 \\
\cmidrule{2-6}          & -1.8  & 918 $\pm$ 30 & 507 $\pm$ 23 & 1821 $\pm$ 43 & 3.27 $\pm$ 0.73 \\
    \bottomrule
    \end{tabular*}%
  \label{table-ratio}%
\end{table*}%

The attenuation factor of on-resonance $\gamma$ rays at ($x$, $\theta_{r}$) can be expressed as
\begin{equation}
  \varepsilon_{ON}(x, \theta_{r})=exp[-(\frac{\mu}{\rho})_{ave}\cdot \rho_{ave}(x, \theta_{r})\cdot L-\sigma_{NRF}\cdot N_{t}(x, \theta_{r})\cdot L ] ,
  \label{eq-on}
\end{equation}
where $({\mu}/{\rho})_{ave}$ is the average mass attenuation coefficient of the CT target (i.e., the interrogated target) on the incident beam path, and $L$ is the diameter of the CT target. $\sigma_{NRF}$ is the NRF reaction cross section of the isotope of interest, and $N_{t}(x, \theta_{r})$ is the isotope number density on the $\gamma$-ray incident path. For the off-resonance $\gamma$ rays, $\sigma_{NRF}$ is negligible; thus, the attenuation factor of the off-resonance $\gamma$ rays is
\begin{equation}
\label{eq-off}
  \varepsilon_{OFF}(x, \theta_{r})=exp[-(\frac{\mu}{\rho})_{ave}\cdot \rho_{ave}(x, \theta_{r})\cdot L].
\end{equation}

The NRF resonant attenuation factor at ($x$,$\theta_{r}$) can be derived as follows~\cite{zen2019demonstration}:
\begin{equation}
	\begin{split}
-ln(\varepsilon_{NRF}) &= -[ln(\varepsilon_{ON})-ln(\varepsilon_{OFF})] \\
&= -[ln(\frac{C_{ON}(x, \theta_{r})}{C_{ON,blank}})-ln(\frac{C_{OFF}(x, \theta_{r})}{C_{OFF,blank}} )] \\
&= \sigma_{NRF}\cdot N_{t}(x, \theta_{r})\cdot L ,
	\end{split}
\label{eq-epsilonNRF}
\end{equation}
where $-ln(\varepsilon_{ON}$) and $-ln(\varepsilon_{OFF}$) are the attenuation factors of the on-resonance and off-resonance $\gamma$ rays, respectively. $C_{ON}(x, \theta_{r})$ and $C_{ON,blank}$ denote the NRF yields recorded by the transmission detectors with and without the CT target, respectively.
$C_{OFF}(x, \theta_{r})$ and $C_{OFF,blank}$ are the integration yields of the spectral region of interest (ROI) recorded by the LaBr$_{3}$(Ce) detector with and without the CT target, respectively.
Note that $\varepsilon_{OFF}$ is an approximate estimate of the atomic attenuation effect of on-resonance $\gamma$ rays in Eq.~\ref{eq-epsilonNRF} when a narrow-band $\gamma$-ray beam is used. Consequently, the NRF attenuation factor depends only on $N_{t}(x, \theta_{r})$, which is required to reconstruct the CT images of SNMs.

\subsection{Simulation algorithm}

To model the NRF process in this study, we developed a new class, G4NRF, in the Geant4 toolkit~\cite{G41,Luo2017}. The pure virtual method G4VUserPhysicslist::ConstructProcess() was implemented in the simulation, and the method AddDiscreteProcess() was used to register the NRF process. Introducing a customized NRF process into the simulation requires the implementation of two features. First, the cross sections for the interaction must be provided; second, the final state resulting from the interaction must be determined. A series of NRF cross sections was calculated using Eq.~\ref{eq4}. Information on the final states was obtained using Eq.~\ref{eq7}. The transitions to the ground states and first excited states of $^{235,238}$U are considered. The HPGe detectors have an energy resolution of 0.1\% (RMS), which can be achieved using present detector technology. The Ge crystals are 10 cm in diameter and 10 cm in length. The full-energy peak efficiency of each HPGe detector was also simulated with the Geant4 toolkit.

\section{Results}
\label{sec-2}
\subsection{Isotope identification by sNRF signature}

\begin{figure}
\centering
\includegraphics[width=7 cm,clip]{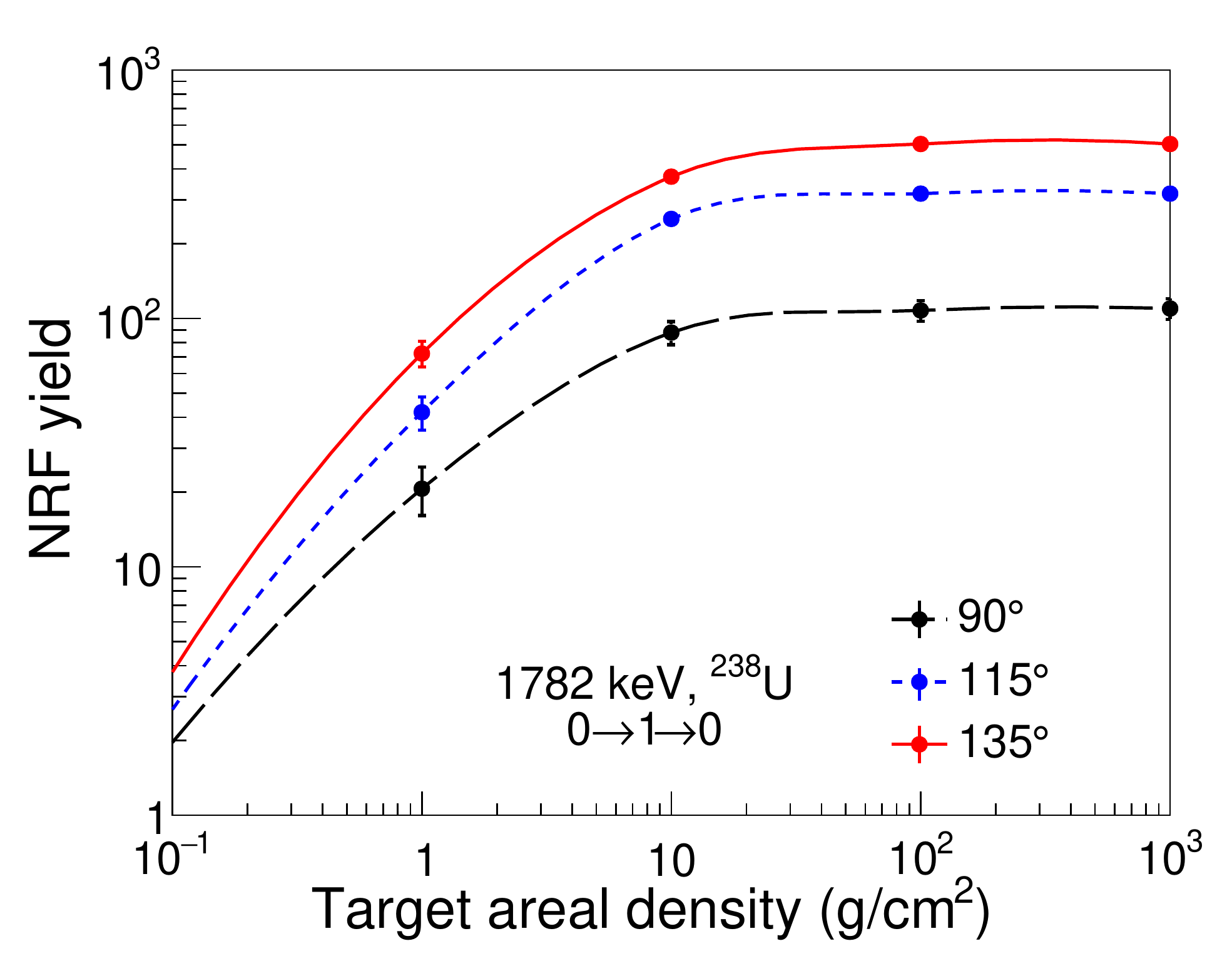}
\caption{\label{f-angle} The 1782-keV NRF yield as a function of target areal density at detection angles of 90$^{\circ}$, 115$^{\circ}$, and 135$^{\circ}$. Only the statistical uncertainty is considered here. The isotopic composition of the uranium rod is 80\% $^{235}$U and 20\% $^{238}$U.}
\end{figure}

\begin{figure*}
\includegraphics[width=12cm,clip]{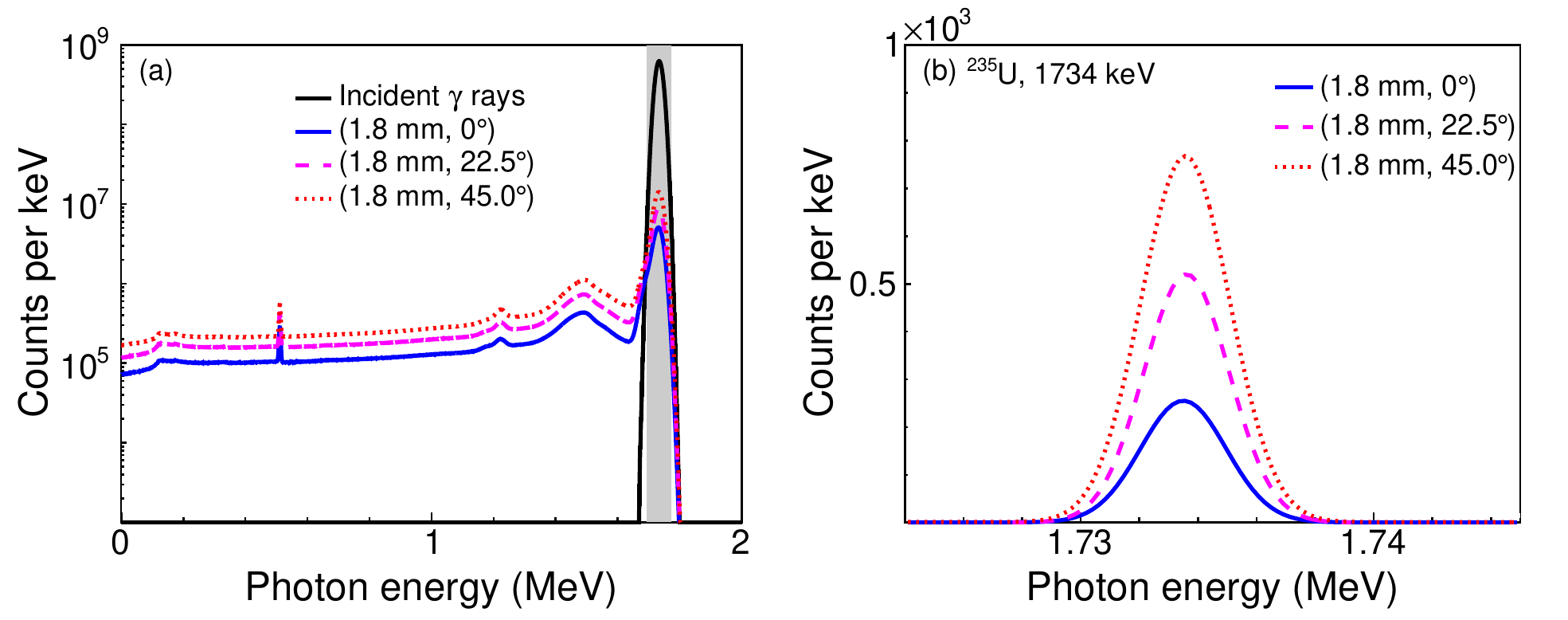}
\caption{\label{f2}(a) $\gamma$-ray spectra detected by the LaBr$_{3}$(Ce) detector at the points (1.8 mm, 0$^{\circ}$) (blue solid line), (1.8 mm, 22.5$^{\circ}$) (magenta dashed line), and (1.8 mm, 45$^{\circ}$) (red dotted line). The energy spectrum of the incident $\gamma$-ray beam (black solid line) is also shown. The shaded area represents the ROI at 1694--1774 keV. (b) Magnified spectra showing the tNRF signal of $^{235}$U around 1734 keV recorded by the transmission detectors when the interrogated object was located at (1.8 mm, 0$^{\circ}$) (blue solid line), (1.8 mm, 22.5$^{\circ}$) (magenta dashed line), and (1.8 mm, 45$^{\circ}$) (red dotted line). }
\end{figure*}

In sNRF spectroscopy, the target is irradiated by a photon beam with a Gaussian energy distribution [centroid energy of 1.76 MeV and energy spread of 3\% in standard deviation (SD)] and a photon intensity of 10$^{10}$ photons per second, which can be readily delivered by a state-of-the-art LCS $\gamma$-ray source~\cite{Weller2015,tanaka2020current}. Among the eight energy spectra obtained in the 1D scan, the sNRF signatures at 1687 keV ($^{235}$U), 1734 keV ($^{235}$U), 1737 keV ($^{238}$U), and 1782 keV ($^{238}$U) appear only in the spectra obtained at $x$ = 1.8 and -1.8 mm. The presence of these sNRF signatures gives a preliminary estimate of the SNM isotopic composition of the interrogated target. This result can potentially reveal a 1D map of SNM isotopes, as reported in the literature~\cite{kikuzawa2009nondestructive,toyokawa2011two}. Moreover, these sNRF signatures can potentially be used for the tomographic imaging of multiple isotopes, which is an interesting issue to study.

Fig.~\ref{f5} shows typical energy spectra of $\gamma$ rays recorded by the scattering detectors at scan points of $x$ = 1.8 and 5.6 mm. The NRF signals are simultaneously observed in the former and disappear in the latter. The NRF $\gamma$-ray peak at 1687 keV is caused by the transition from the resonant state of $^{235}$U at 1734 keV to the 9/2$^{-}$ excited state at 46 keV. The peak near 1734 keV is caused by the transition from the 1734 keV level in $^{235}$U to the ground state and the transition from the 1782 keV state in $^{238}$U to the first excited state at 45 keV (with photon emission at 1737 keV). Note that these two closely spaced NRF lines cannot be well discriminated owing to spectral broadening resulting from imperfect detector resolution. The NRF peaks at 1687, 1734 (or 1737), and 1782 keV are then fitted with four Gaussian distributions on top of an exponentially decaying continuum background. The fitting function for these NRF peaks is written as
\begin{equation}
f(E) = exp(c_{1} + c_{2} E) + \sum_{k=1}^{4}\frac{a_{k}}{\sqrt{2\pi} \sigma_{k}} exp\left[-\frac{(E-E_{k})^2}{2 \sigma_{k}^{2}}\right],
\label{eq-Fit}
\end{equation}
where $c_{1}$ and $c_{2}$ describe the shape of the background, and $a_{k}$, $E_{k}$, and $\sigma_{k}$ are the area, mean, and SD fit parameters of the $k^{th}$ peak. The fitting curve is shown in Fig.~\ref{f5} (b). The corresponding NRF yields, $Y_{1687}$, $Y_{1734,1737}$, and $Y_{1782}$, were obtained. Because the branching ratio (denoted as $b_{1}$) of the 1734 keV transition to the 1687 keV transition is 100:60(20), the NRF yield for the 1734 keV transition was further deduced as $Y_{1734}^{ded} = b_{1} Y_{1687}$. Similarly, the yield for the 1737 keV transition, $Y_{1737}^{ded}$, was also deduced. Moreover, the NRF yield for the overlapping peak near 1734 keV was estimated as $Y_{1734,1737}^{ded}$=$Y_{1734}^{ded}+Y_{1737}^{ded}$, as shown in Table~\ref{table-NRFyilds}. The extracted $Y_{1734,1737}^{ded}$ values agree well with $Y_{1734,1737}$, indicating that the effect of branching ratios was implemented appropriately in our simulations. The significance of the sNRF signals of $^{235}$U can be expressed in units of $\sigma$ as $\Delta_{sNRF}$ = $S/{\delta}S$, where $S$ is the NRF peak yield, and ${\delta}S$ is the corresponding statistical error. The result shows that a peak significance of $\Delta_{sNRF}^{0}$ = 48.9$\sigma$ can be obtained within an sNRF scan time of $t_{sNRF}^{0}$ = 8 s.

The isotope ratio of $^{235}$U to $^{238}$U, ${^{235}U}/{^{238}U}$, is related to the NRF yield ratio, $^{235}Y/^{238}Y$, by the following equation:
\begin{equation}
  \frac{^{235}U}{^{238}U}=\frac{^{235}Y}{^{238}Y}\frac{I(E_{^{238}U})}{I(E_{^{235}U})}\frac{W_{^{238}U}(\theta)}{W_{^{235}U}(\theta)}\frac{\varepsilon_{^{238}U}}{\varepsilon_{^{235}U}}\frac{br_{^{238}U}}{br_{^{235}U}}\frac{\int\sigma_{^{238}U}(E)dE}{\int\sigma_{^{235}U}(E)dE},
  \label{eq-ratio}
\end{equation}
where ${^{235}Y}/{^{238}Y}$ is the peak yield ratio, and ${I(E_{^{238}U})}/{I(E_{^{235}U})}$ is the ratio of the incident $\gamma$-ray intensity. ${\varepsilon_{^{238}U}}/{\varepsilon_{^{235}U}}$ is the detection efficiency ratio of the HPGe detector calculated using the MCLCSS code~\cite{luo20114d} in Geant4. ${br_{^{238}U}}/{br_{^{235}U}}$ is the ratio of the absolute branching ratios. ${\int\sigma_{^{238}U}(E)dE}/{\int\sigma_{^{235}U}(E)dE}$ is given by Eq.~\ref{eq4}. ${W_{^{238}U}(\theta)}/{W_{^{235}U}(\theta)}$ is the angular momentum ratio of $^{238}$U and $^{235}$U NRF emission~\cite{Fagg1959}.
According to Eq.~\ref{eq-ratio}, the abundance ratios of ${^{235}U}/{^{238}U}$ at $x$ = 1.8 and -1.8 mm for three $^{235}$U enrichments were calculated on the basis of the NRF yields at 1687 and 1782 keV, as shown in Table~\ref{table-ratio}. The predicted isotope ratios are consistent with the theoretical values within the uncertainty. Note, however, that the attenuation of $\gamma$ rays with different energies as they penetrate the wrapping materials should be considered in Eq.~\ref{eq-ratio} to improve the effectiveness of the isotope ratio prediction for thicker shielding.

\begin{figure*}
\includegraphics[width=12cm,clip]{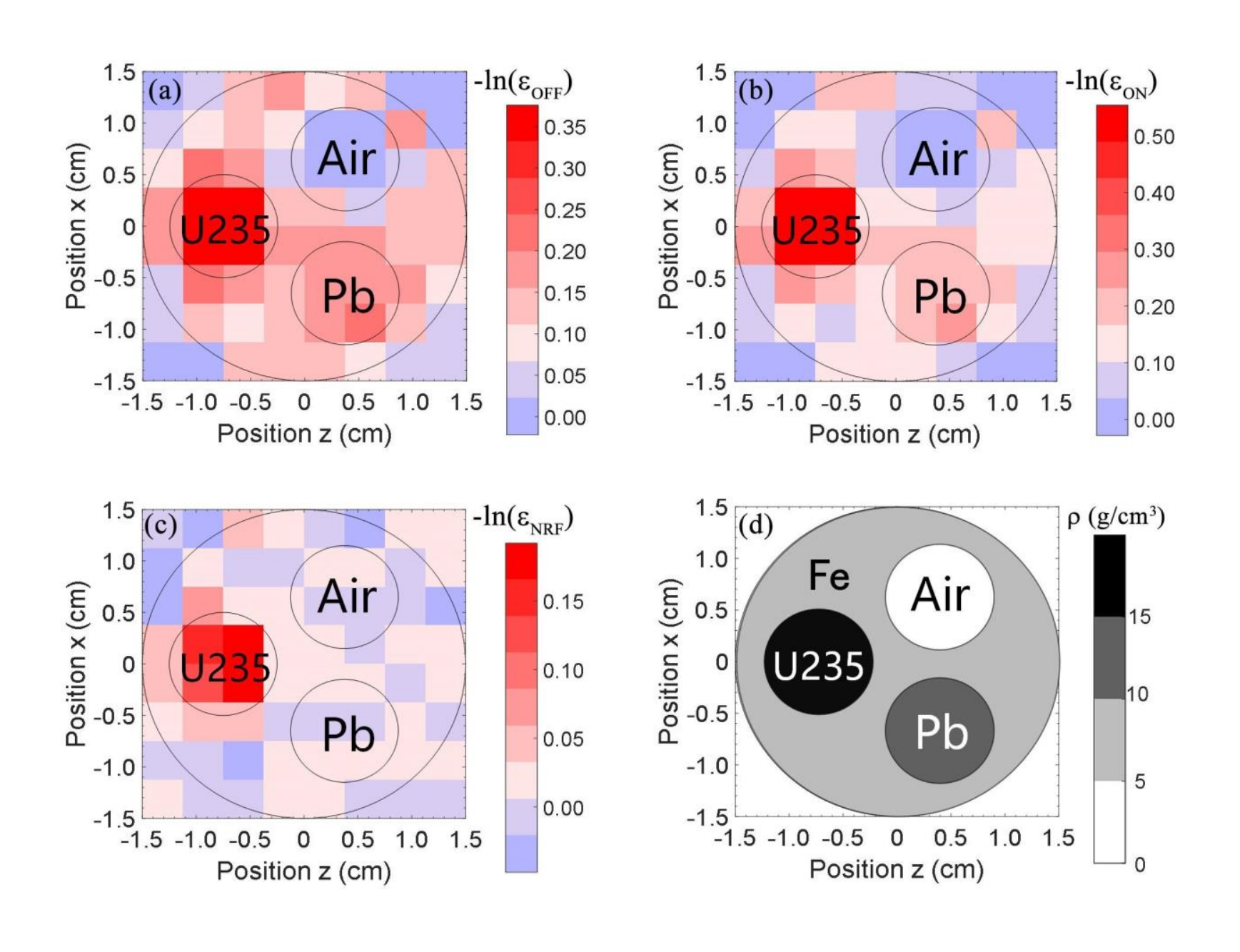}
\caption{\label{f3}(a) Image of non-resonant atomic attenuation factor $-ln(\varepsilon_{OFF})$. (b) Image of on-resonant attenuation factor $?ln(\varepsilon_{ON})$. (c) Image of nuclear resonance attenuation factor $-ln(\varepsilon_{NRF})$.  (d) Geometry and density of the interrogated target.}
\end{figure*}

\begin{figure}
\centering
\includegraphics[width=8cm,clip]{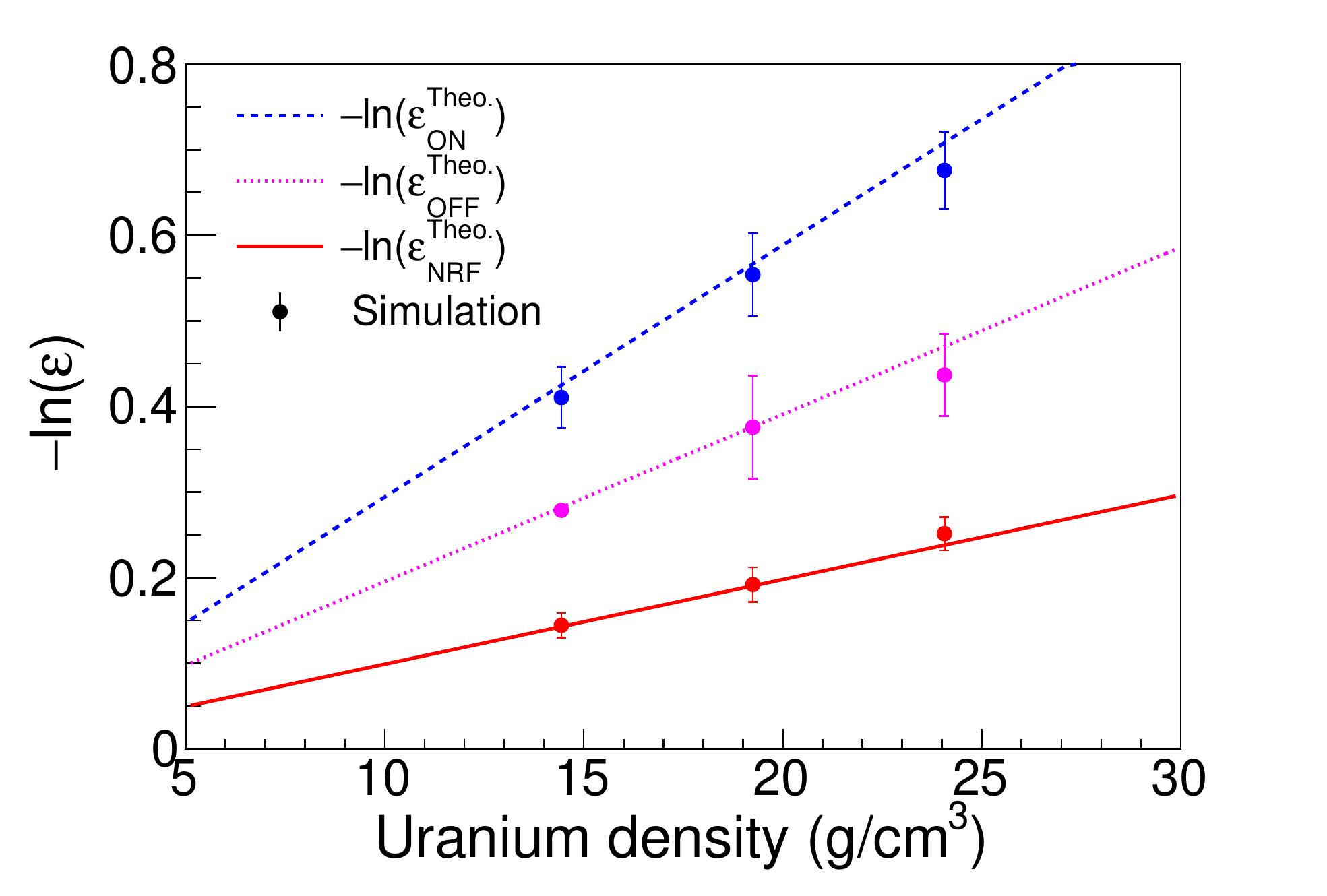}
\caption{\label{fepsilon}Dependence of attenuation factors $-ln(\varepsilon_{ON})$, $-ln(\varepsilon_{OFF})$, and $-ln(\varepsilon_{NRF})$ on the $^{235}$U density. The theoretical predictions given by Eqs.~\ref{eq-on}, \ref{eq-off}, and \ref{eq-epsilonNRF} are also shown for comparison.}
\end{figure}

Fig.~\ref{f-angle} shows the simulated NRF yields of the 1782 keV line when the HPGe detectors are located at angles of 90$^{\circ}$, 115$^{\circ}$, and 135$^{\circ}$. It is shown that the NRF yields increase with detection angle because the $W(\theta)$ value for the 1782 keV line at 135$^{\circ}$ is larger than those at the other two angles.
Consequently, a detection angle of 135$^{\circ}$ is employed in our study (see Fig.~\ref{f1}).

\subsection{Tomographic imaging using tNRF signature}

In tNRF imaging, the uranium rod is ideally assumed to be composed of pure $^{235}$U, and its default density is set to 19 g/cm$^{3}$ to reduce the computational requirements. The interrogating $\gamma$-ray beam has a Gaussian distribution (a centroid energy of 1734 keV and an energy spread of 1\% in SD) and a photon intensity of 10$^{10}$ photons per second. Fig.~\ref{f2} shows typical spectra obtained by the transmission detectors and LaBr$_{3}$(Ce) detectors at measurement points of (1.8 mm, 0$^{\circ}$), (1.8 mm, 22.5$^{\circ}$), and (1.8 mm, 45$^{\circ}$), where the uranium rod, lead rod, and air rod, respectively, are in the path of the interrogating $\gamma$-ray beam.
As shown in Fig.~\ref{f2} (a), the energy spectra recorded by the LaBr$_{3}$(Ce) detector are different because they depend on the atomic attenuation coefficients of the penetrated materials.
The 1734 keV peak intensity recorded by the HPGe detectors at the scan point (1.8 mm, 0$^{\circ}$) is significantly lower than those at the other two scan points. The reason is that the intensity of the $\gamma$-ray beam transmitted through the CT target decreases with $\theta_{r}$ owing to strong resonant absorption.

As mentioned above, 65 sets of spectra were obtained.
On the basis of the NRF peak yields at 1687 and 1734 keV, the $\epsilon_{ON}(x, \theta_{r})$ values for these scan points are obtained. Similarly, the values of $\epsilon_{OFF}(x, \theta_{r})$ are obtained on the basis of the integration of the spectral ROI recorded by the LaBr$_{3}$(Ce) detector [see Fig.~\ref{f2} (a)].
These extracted values are further incorporated into the simultaneous algebraic reconstruction techniques (SART) algorithm~\cite{andersen1984simultaneous}, which is suitable for the reconstruction of high-quality images with limited observation angles.
Fig.~\ref{f3} shows images of $-ln(\varepsilon_{OFF})$, $-ln(\varepsilon_{ON})$, and $-ln(\varepsilon_{NRF})$ reconstructed by SART. The $-ln(\varepsilon_{OFF})$ values decrease with the density of the materials; this behavior is similar to that of a conventional X-ray CT image. In addition, the signal of the uranium rod is clearly enhanced in the $-ln(\varepsilon_{ON})$ image, although the lead rod and air rod are still visible. In the $-ln(\varepsilon_{NRF})$ image, the signals produced by the air and lead rod do not appear, and the contrast of the uranium rod is higher than that in the $-ln(\varepsilon_{ON})$ image. These results demonstrate that the tNRF imaging method can be used to spatially discriminate a suspicious SNM isotope.
The significance of the tNRF signals of $^{235}$U can be expressed in units of $\sigma$ as $\Delta_{tNRF}$ = $S/{\delta}S$, where $S$ and ${\delta}S$ are the average $|-ln(\varepsilon_{NRF})|$ value of four pixels in the uranium rod region and that of the remaining pixels, respectively. A tNRF imaging time of $t_{tNRF}^{0}$ 65 s is expected to yield a significance of $\Delta_{tNRF}^{0}$ = 8.0$\sigma$. This relatively long time is correlated with the photon flux, SNM concentration, and weak response of tNRF imaging to SNMs.

\newcommand{\tabincell}[2]{\begin{tabular}{@{}#1@{}}#2\end{tabular}}
\begin{table*}
\caption{ Resonant level, NRF $\gamma$-ray energy, $\sigma_{int}$, $ln(\varepsilon_{NRF})$, and imaging feasibility of several SNMs.}
\begin{tabular*}{120mm}{@{\extracolsep{\fill}}cccccccc}
\toprule
Isotope & $E_{r}$ (keV) & $E_{\gamma}$ (keV) & $I_{\gamma}$ (\%) & $\sigma_{int}$ (eV$\cdot$b) &$ln(\varepsilon_{NRF})$ &Ratio &Imaging Feasibility \\
\midrule
$^{235}$U	 &1734 & \tabincell{c}{1687 \\ 1734} & \tabincell{c}{60(20) \\ 100}	&21.7	&1.88	&1.0	    &feasible \\
$^{238}$U	 &2468 & 2468 & 100    &90.9	&5.59	&3.0	&feasible \\
$^{239}$Pu   &2040	& 2040 & 100 &8.0	&0.60	&0.3	&questionable \\
$^{240}$Pu   &2152	& 2152 & 100 &34.4	&2.46	&1.3	&feasible \\
$^{237}$Np   &1729	& 1729 & 100 &10.6	&0.92	&0.5	&questionable \\
\bottomrule
\label{t1}
\end{tabular*}
\end{table*}

To evaluate the dependence of the attenuation factors on SNM density, we reconstructed CT images of $^{235}$U at artificial target densities of 14, 19, and 24 cm$^{3}$. Reconstructed images similar to those in Fig.~\ref{f3} were obtained. The average $-ln(\varepsilon_{ON})$, $-ln(\varepsilon_{OFF})$, and $-ln(\varepsilon_{NRF})$ values over the uranium rod region were then extracted, as shown in Fig.~\ref{fepsilon}. One can see that the extracted values increase with increasing $^{235}$U density, which is consistent with the theoretical predictions. Subsequently, using Eq.~\ref{eq-epsilonNRF}, we also calculated the $-ln(\varepsilon_{NRF})$ values of tNRF images of the SNMs  $^{238}$U, $^{239, 240}$Pu, and $^{237}$Np (see Table~\ref{t1}). The expected $-ln(\varepsilon_{NRF})$ values for $^{238}$U (2468 keV) and $^{240}$Pu (2152 keV) are significantly larger than that for $^{235}$U (1734 keV), indicating excellent potential for the use of tNRF-CT for these SNMs. However, the $-ln(\varepsilon_{NRF})$ values for $^{239}$Pu and $^{237}$Np are smaller than that for $^{235}$U.
This result suggests that the interrogations of both $^{239}$Pu and $^{237}$Np are questionable; a higher beam flux or longer imaging time would be required to obtain better results.

\section{Discussion}
\label{sec-discussion}
In SNM interrogation, two errors can occur: a false alarm, where the test indicates that the SNM is present when in fact the ``all clear'' hypothesis is correct, and a missed detection, where the test shows ``all clear'' but the SNM is in fact present. We attempt to demonstrate the scientific justification of our proposed method by comparing it with the use of the tNRF method alone in the context of balancing the measurement time and the missed-detection rate in an SNM interrogation.

\begin{figure}
\centering
\includegraphics[width=9cm,clip]{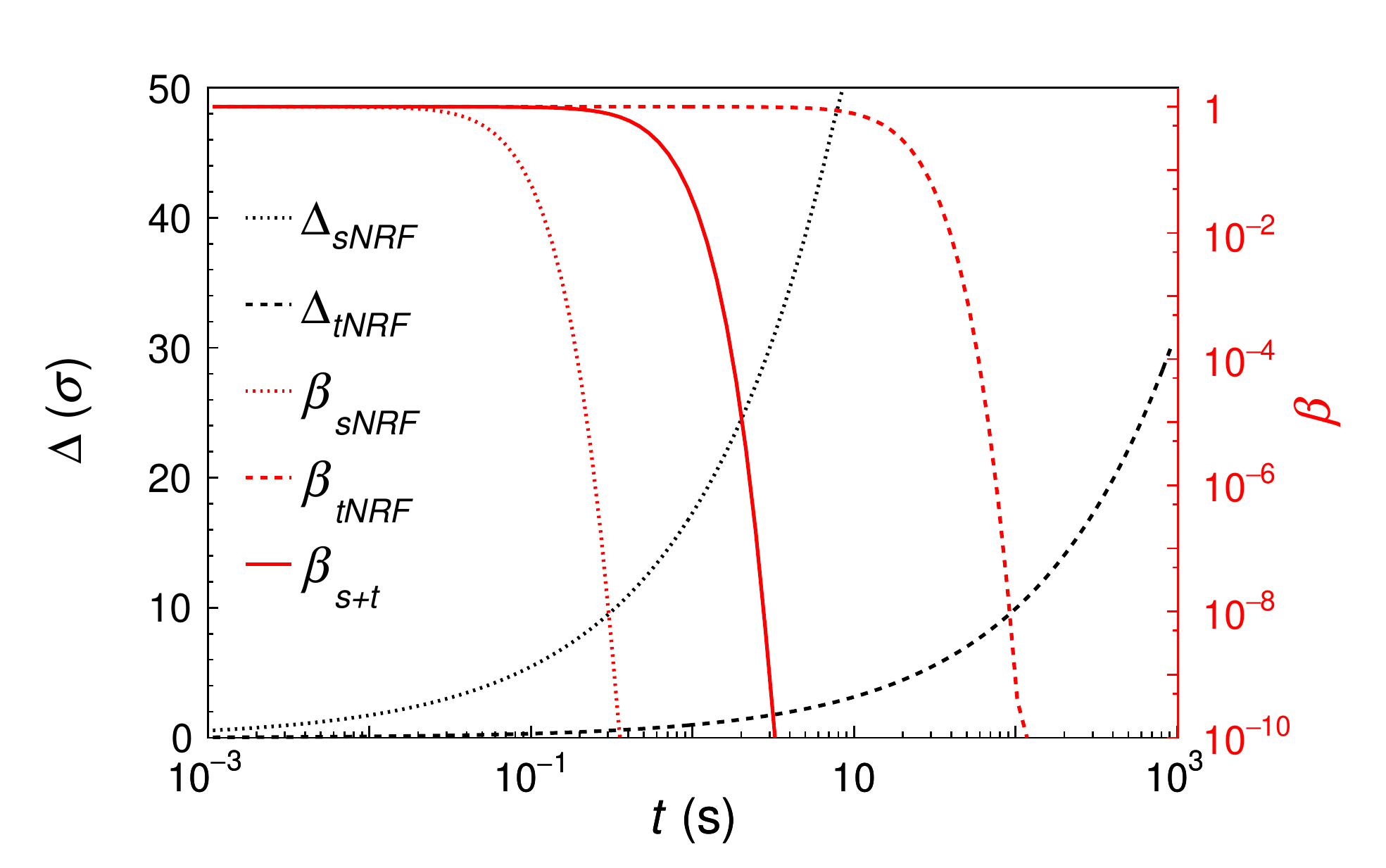}
\caption{\label{missed-detection}Variation of $\Delta$ and $\beta$ for the tNRF method alone (dashed line), the sNRF method alone (dotted line), and sNRF plus tNRF (solid line) for interrogation time $t$. The alarm threshold of $\Delta_{th}$ = 3.9$\sigma$ is set for the calculations of $\beta_{sNRF}$, $\beta_{tNRF}$, and $\beta_{t+s}$ to achieve a false-alarm rate of 10$^{-4}$. The beam intensity is 10$^{10}$ photons per second.}
\end{figure}

 $\Delta_{sNRF}$ as a function of sNRF scan time $t_{sNRF}$ can be expressed as $\Delta_{sNRF}$ = $\Delta_{sNRF}^{0}\times\sqrt{t_{sNRF}/t_{sNRF}^{0}}$, considering only the statistical fluctuation because the background is negligible. Similarly, $\Delta_{tNRF}$ for a tNRF imaging time $t_{tNRF}$ can be given as $\Delta_{tNRF}$ = $\Delta_{tNRF}^{0}\times\sqrt{t_{tNRF}/t_{tNRF}^{0}}$.
To determine whether an interrogated object contains $^{235}$U, the decision rule is to alarm if $\Delta_{sNRF}$ or $\Delta_{tNRF}$ exceeds a test threshold $\Delta_{th}$. The missed-detection rates $\beta$ are easily obtained using $\Delta_{sNRF}$ and $\Delta_{tNRF}$ as $\beta_{sNRF}$ = $\Phi(\Delta_{th}-\Delta_{sNRF})$ and $\beta_{tNRF}$ = $\Phi(\Delta_{th}-\Delta_{tNRF})$. Here $\Phi$ is the cumulative distribution function of a normal distribution centered at zero with variance unity~\cite{zkproof1}. Here, because the sNRF and tNRF methods are both used, the missed-detection rate can be bounded as
\begin{equation}
	\begin{split}
 \beta_{s+t} &= \Phi(\Delta_{th}-\Delta_{sNRF})\times\Phi(\Delta_{th}-\Delta_{tNRF}) \\
&= \Phi(\Delta_{th}-\Delta_{sNRF}^{0}\times\sqrt{t_{s+t}/t^{0}}) \\
&\times\Phi(\Delta_{th}-\Delta_{tNRF}^{0}\times\sqrt{t_{s+t}/t^{0}}),
   \end{split}
\end{equation}
where $t_{s+t}$ is the total interrogation time for both sNRF scanning and tNRF imaging. In addition,  $t^{0}$ is given by $t^{0}$ = $t_{tNRF}^{0}$ + $t_{sNRF}^{0}$.

 Fig.~\ref{missed-detection} shows the expected $\beta$ as a function of time when the tNRF method alone, the sNRF method alone, and sNRF plus tNRF are used. When only tNRF is used, an interrogation time of $\sim$53 s is required to reach $\beta$ = 10$^{-3}$. This result indicates that all objects containing $^{235}$U can be detected with greater than 99.9\% probability in a 53 s interrogation. In practice, an important issue would be to achieve a low missed-detection rate with a shorter interrogation time. In addition, although the use of the sNRF method alone does not afford imaging capability, it requires much less time to reach the same $\beta$ value. Thus, the combination of sNRF scanning and tNRF imaging is considered in our study to address this shortcoming. This combination yields a missed-detection rate of 10$^{-3}$ within an interrogation time of 1.5 s, which is one order of magnitude lower than that when only tNRF imaging is used.

  We performed additional simulations to evaluate the influence of $W(\theta)$ on the sNRF yields because the excitation and de-excitation of the 1734 keV state of $^{235}$U are affected by spin selection uncertainty and mixing ratio unavailability. In the simulations, the non-uniform distributions of $W(\theta)$ obtained from the transition sequences 7/2 $\rightarrow$ 9/2 $\rightarrow$ 7/2, q-q and 7/2 $\rightarrow$ 9/2 $\rightarrow$9/2, q-q (see Fig.~\ref{angular}) are applied. It is found that the NRF yield $Y_{1687}$ increases by a factor of $\sim$1.1, and $Y_{1734,1737}$ decreases by a factor of $\sim$0.9. Thus, the extracted NRF yields do not differ significantly from those obtained considering an isotropic $W(\theta)$.

\section{Conclusion}

The interrogation of SNMs is an essential technique to prevent global nuclear proliferation. In this work, we combined sNRF and tNRF spectroscopy to achieve the rapid identification and tomographic imaging of SNMs. It was shown that the isotopic composition of $^{235, 238}$U and their isotope ratio can be determined from the photon emission of the resonant states at 1734 and 1782 keV using sNRF scanning. The spatial distribution of $^{235}$U concealed in a 3-cm-diameter iron rod can be well visualized using tNRF imaging. We conclude that the combination of sNRF scanning with tNRF imaging has the advantage of achieving a significantly lower missed-detection rate within a realistic interrogation time compared to that obtained using only tNRF spectroscopy.

\end{document}